\documentclass[preprints,article,accept,moreauthors,pdftex,10pt,a4paper]{Definitions/mdpi} 

\usepackage{amsmath,amssymb}
\usepackage[normalem]{ulem}

\firstpage{1} 
\pubvolume{xx}
\issuenum{1}
\articlenumber{5}
\pubyear{2018}
\copyrightyear{2018}
\history{Received: date; Accepted: date; Published: date}

\Title{Many-body physics of low-density dipolar bosons in box potentials}

\Author{Tommaso Macr\`i $^{1}$ and Fabio Cinti $^{2}$}

\AuthorNames{Firstname Lastname, Firstname Lastname and Firstname Lastname}

\address{
$^{1}$ \quad Departamento de F\'isica Te\'orica e Experimental, Universidade Federal do Rio Grande do Norte, and International Institute of Physics, Natal-RN, Brazil; macri@fisica.ufrn.br\\
$^{2}$ \quad Department of Pure and Applied Mathematics, University of Johannesburg, Johannesburg, South Africa; fcinti@uj.ac.za}

\abstract{Crystallization is a generic phenomenon in classical and quantum mechanics arising in a variety of physical systems. In this work we focus on a specific platform, ultracold dipolar bosons, which can be realized in experiments with dilute gases.
We review the relevant ingredients leading to crystallization, namely the interplay of contact and dipole-dipole interactions and system density, as well as the numerical algorithm employed. We characterize the many-body phases investigating correlations and superfluidity.}


\begin{document}


\section{Introduction}
 
Cluster phases are ubiquitous in diverse systems, from biological structures \cite{Butenko2009} to soft matter \cite{Likos2001,PhysRevLett.118.067001} to ultra-cold atomic and molecular condensates \cite{Cinti:2014aa}. 
In many respects, particle aggregates display fascinating properties, whose features can be mainly described microscopically by means of effective two-body potentials.
In a simple mean field picture, a cluster crystalline phase (or, more formally, a spontaneous breaking of translational symmetry)
appears when the Fourier transform (when it exists) of the particle-particle interaction exhibits a negative region with a local minima. 
The cluster phase has then a density modulation with wavelength corresponding to those minima. 
The validity of this condition has been investigated also in the quantum regime, and it holds also if one considers fermionic or bosonic quantum gases \cite{Cinti2010b}. In particular, 
having potentials displaying a soft-shoulder-like shape at short distance is a crucial  ingredient for designing 
novel quantum phases. As an example, alkaline atoms off-resonantly excited to 
Rydberg states 
display an effective interaction with a soft-core at short distances 
that causes a supersolid phase \cite{Henkel2011,Macri:2014aa,PhysRevA.87.061602,1367-2630-16-3-033038}. 
A supersolid is a phase of matter that simultaneously accommodates diagonal as well as off-diagonal long-range order, which means that particles self-assemble into a rigid, regular crystal but at the same time can they can rotate with a corresponding reduced moment of inertia, which is an indication of the presence of a finite superfluid fraction
\cite{Cinti2010b,PhysRevLett.108.175301,Saccani2011,Boninsegni2012,PhysRevLett.113.240407,RevModPhys.84.759}

Recently, ground-breaking experiments with bosonic dipolar condensates 
(Dysprosium and Erbium)
demonstrated the existence of \textit{self-bound droplets} in trapped configurations as well as in free space \cite{Kadau2016,PhysRevX.6.041039}. 
These experiments provides a useful isolated system to probe general quantum-mechanical properties related to  cluster phases. 
A mean-field treatment mainly based on a generalized nonlocal nonlinear 
Schr\"odinger equation reveal a good agreement 
with the density distribution and the excitation spectra observed in the laboratory \cite{PhysRevA.93.061603}. 
The same methodology has been successfully implemented to Bose-Bose mixtures \cite{PhysRevA.97.053623,Cappellaro2017}.
In addition, quantum Monte Carlo simulations (QMC) have investigated the zero temperature phase diagram of bosons interacting via dipolar interactions in three dimensions in free space \cite{PhysRevLett.119.215302}.
Other theoretical works have shown how the physical nature of self-bound droplets smoothly evolves from classical to quantum mechanical as the range of the repulsive two-body potential increases \cite{PhysRevA.96.013627,PhysRevA.98.023618}.

In the present paper, by means of first-principles numerical simulations, 
we characterize the many-body physics of bosons interacting via anisotropic dipole-dipole potentials with 
different densities and interaction strengths. In particular we carry out simulation to better understand 
the quantum behaviour of the system that  
supports recent experimental findings about the stability of droplets. 

The rest of the paper is organized as follows: In section \ref{section2} we introduce the microscopic potential that describes the physics of dipolar systems in three-dimension, whereas in section \ref{section3}
we present the details of the many-body properties known in literature.In section \ref{section4}, we illustrate our results, outlining our conclusions in section \ref{section5}.

 \section{Two-body physics of dipole-dipole interactions}\label{section2}

The two body-potential describing a short-range and an anisotropic long-range dipolar interaction reads 
\begin{equation} \label{v1}
V({\bf r}) = \frac{4\pi\hbar^2}{m}\,a_s\, \delta({\bf r})\,+\,\frac{C_{dd}}{4\pi}\, \frac{\left(1-3\cos^2\theta\right)}{r^3}
\end{equation}
where $m$ is the particle mass, while $a_s$ is the s-wave scattering length and $a_d=mC_{dd}/4\pi\hbar^2$ the dipolar length representing  the characteristic length of the dipole interactions $C_{dd}/4\pi$ \cite{Lahaye2009}. 
$\theta$ denotes the angle between the vector ${\bf r}$ and the polarization axis of the dipoles, the $z$-axis.
Equation \eqref{v1} applies both to vertically aligned dipoles interacting via magnetic or electric dipole moments \cite{Lahaye2009}.
The pseudo-potential  $V({\bf r})$ has been already discussed by Yi and Yu  \cite{Yi2000,PhysRevA.63.053607} making use of the Born approximation. More realistic model potentials, that include also a Van der Walls short range potentials, were recently considered in \cite{PhysRevA.94.063638} for a more detailed 
description of Dy two-body collisions. 
These potentials eventually lead to the same effective potential of \cite{Yi2000,PhysRevA.63.053607} , with a dipolar length depending on the short range scattering length.

\begin{figure}[t]
\centering
\includegraphics[width=10cm]{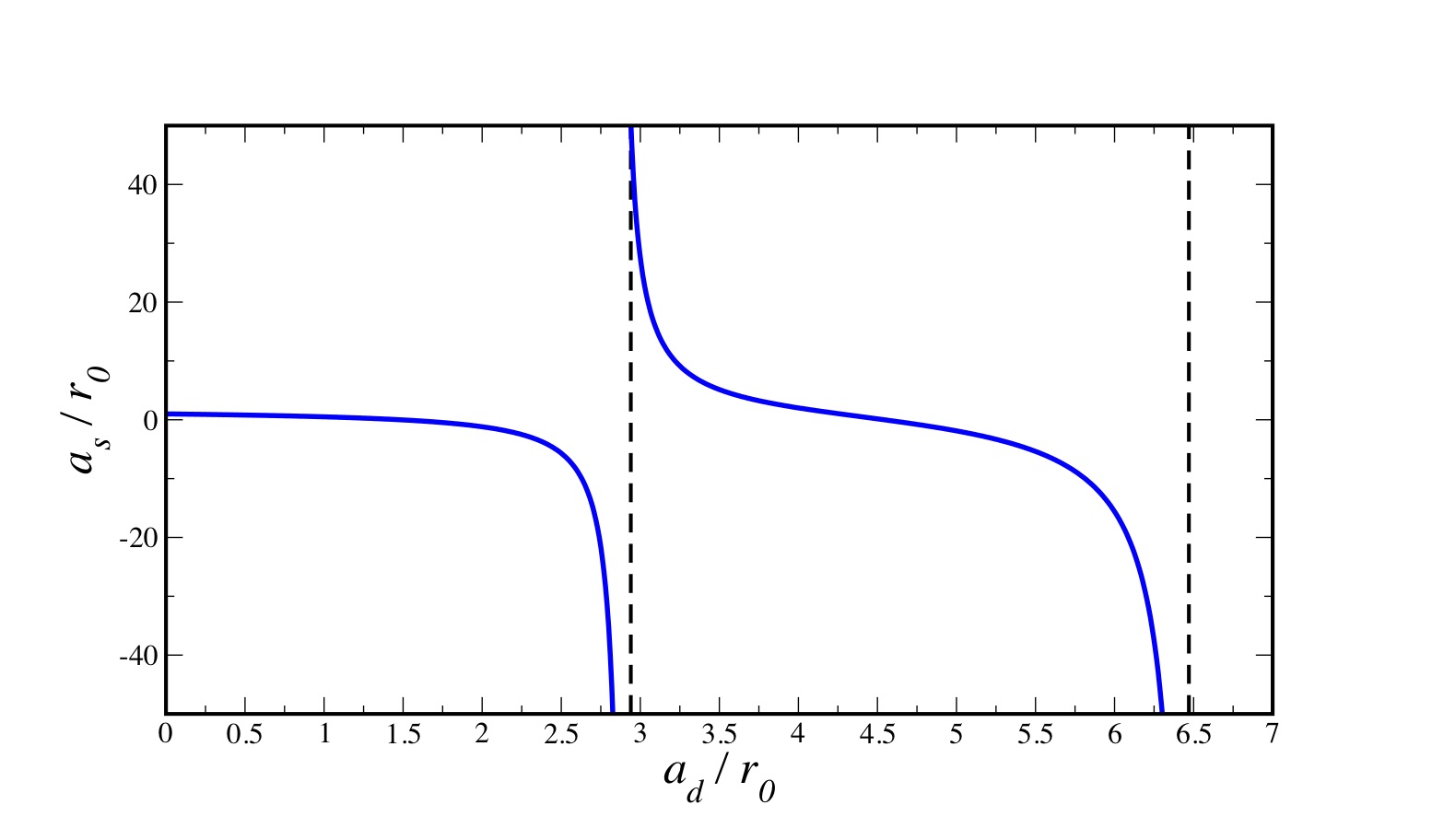}
\caption{Relation between scattering length $a_s$ and dipolar length $a_d$ in units of $r_0$, see text.}\label{figure1}
\end{figure}   

In our first-principle simulations based on a QMC technique (see section  \ref{section3}), $a_d$ as well as $a_s$ are defined in terms of a unit length $r_0$, $\varepsilon_0=\hbar^2/mr^2_0$ being the corresponding energy with the effective pairwise potential ${\cal V}({\bf r})$:
\begin{equation} \label{v2}
{\cal V}({\bf r})=
\left\{
\begin{array}{ll}
\infty & \text{if}~r<r_0,\\
\\
\frac{C_{dd}}{4\pi}\,\frac{1-3\cos^2\theta}{r^3} & \text{if}~r\ge r_0.
\end{array}
\right.
\end{equation}
${\cal V}({\bf r})$ describes a short-range hard core with cut-off at $r_0$ and the usual anisotropic dipolar component. 
The authors of \cite{PhysRevLett.97.160402,Ronen2007} have computed numerically 
the full low-energy scattering amplitude 
for the same model potential in equation (\ref{v2}). 
A  comparison with equation (\ref{v1}) found a quite good agreement.

As a useful example, we now consider the Dysprosium's isotopes $^{162}$Dy ($a_d = 129.2\, a_0$, $a_0$ being the Bohr radius) and $^{164}$Dy ($a_d=130.8\, a_0$), relevant to the experimental observations by Kadau et al. \cite{Kadau2016}. In addition, the background scattering lengths were recently measured for both isotopes (6-8):
$a_s = 122\, a_0$ and $a_s = 92 a_0$, for $^{162}$Dy and $^{164}$Dy, respectively. 

\begin{figure}[t]
\centering
\includegraphics[width=12 cm]{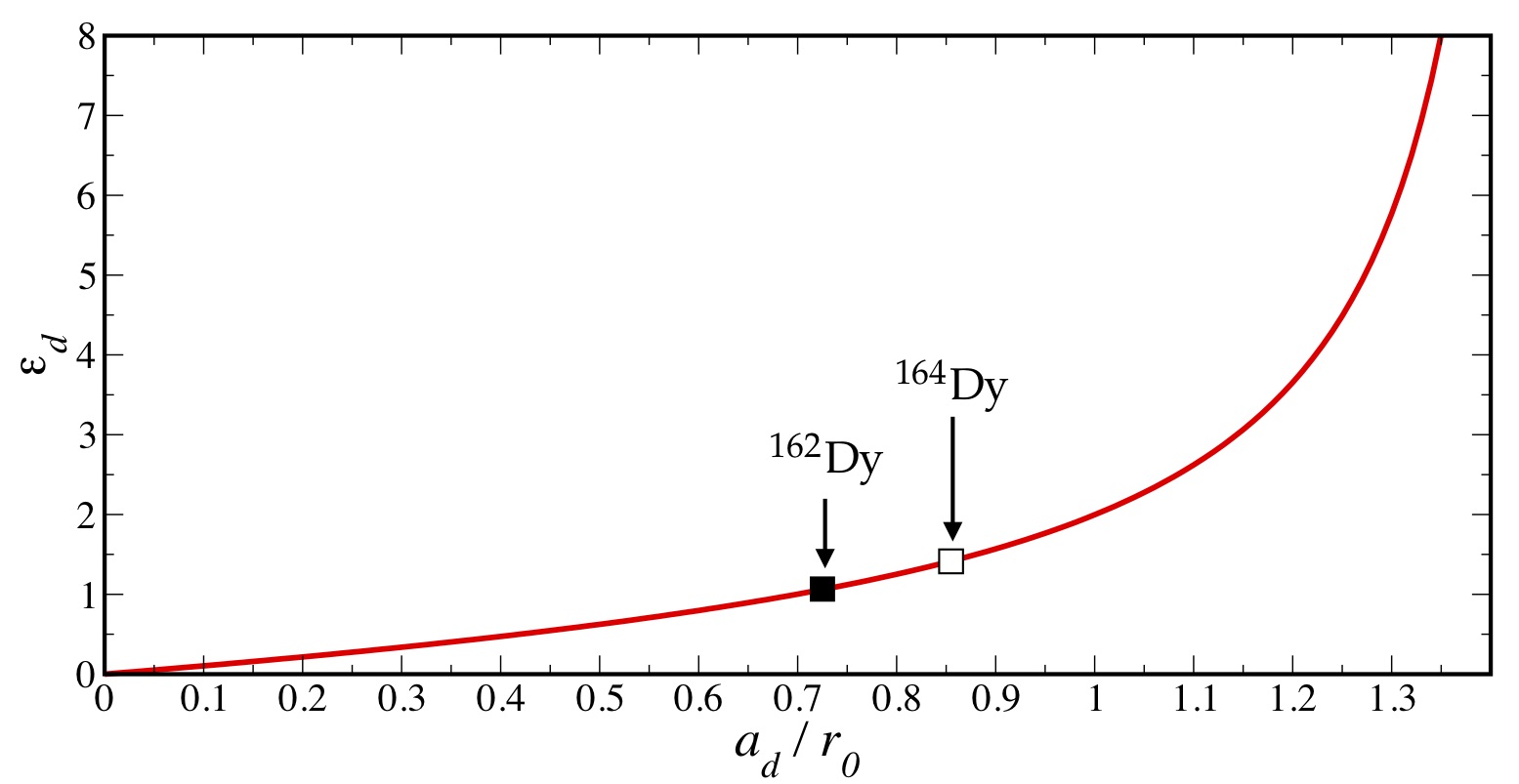}
\caption{Ratio $\epsilon_d=a_d/a_s$ vs $a_d/r_0$. filled and open square and correspond
to experimental parameters for $^{162}$Dy and $^{164}$Dy, respectively, with their background scattering length 
($122 a_0$ and $92 a_0$, respectively).}\label{figure2}
\end{figure}   

In \cite{PhysRevLett.97.160402} the authors extracted the relation between the dipolar length $a_d$ and 
the scattering length $a_s$ for the model potential ${\cal V}({\bf r})$ as shown in fig. \ref{figure1}. 
For $a_d/r_0\approx 2.9$ and $a_d/r_0\approx 6.5$ we observe two resonances at which the scattering length $a_s$ diverges. Moreover, one also extracts the value of $\epsilon_d = a_d/a_s$ as a function of the corresponding ratios $a_s/r_0$ and $a_d/r_0$ (see Fig. \ref{figure2}). From the calculation of $\epsilon_d$ for the two isotopes we can also compute the value of the parameter $r_0$. In particular, for $^{162}$Dy we have $r_0 = 177 a_0$, while considering $^{164}$ Dy 
we get  $r_0 = 154 a_0$.

\section{Many-body properties} \label{section3}

The present section is devoted to discuss many-body properties of bosons interacting via a dipolar interaction. 
The quantum-mechanical Hamiltonian of an ensemble of $N$ interacting identical bosons reads 
\begin{equation} \label{ham}
{\cal H} = -\frac{1}{2}\sum_{i=1}^N \nabla_i^2 + 
 \sum_{i<j}^N {\cal V}(\mathbf{r}_i-\mathbf{r}_j). 
\end{equation}
The Hamiltonian is reported in units of $r_0$ and $\varepsilon_0$. It is important to note that  
the zero-temperature physics is exclusively controlled by the dimensionless interaction strength ${\cal V}_0=mC_{dd}/4\pi\hbar^2r_0$
and  the dimensionless density $nr^3_0$. Recently the zero-temperature phase diagram of 
equation (\ref{ham}) has been throughly investigated applying QMC techniques \cite{PhysRevLett.119.215302}. 
In particular, considering low ${\cal V}_0$ (that is $\lesssim 2.1 $ in the present units),
simulations agree with the mean-field phase boundary predicted by standard Bogoliubov analysis.  
Upon increasing the dipole-dipole interaction we can distinguish two different phases. 
At low densities ($nr^3_0\lesssim5\cdot10^{-3}$) the system displays a cluster phase with vanishing superfluidity 
and characterized by droplet structures with few particles. For higher densities, one observes a phase  marked by elongated filaments with an anisotropic superfluid fraction;
that is, whereas the superfluidity is observed along the direction of the filaments, it results greatly suppressed on the corresponding orthogonal plane, 
excluding then the occurrence of a global supersolid phase. In addition, finite-temperature calculations confirmed the stability of the filament phase against thermal fluctuations and provide an estimate of the superfluid fraction in the weak coupling limit in the framework of the Landau two-fluid model. 
It is important to stress that both cluster phase and filament phase extend from a small positive induced contact potential ($\epsilon_d\gtrsim 1$), corresponding to the experimentally relevant regime of \cite{Kadau2016},
to the strongly coupled limit of large dipolar interactions.

Theoretical results obtained solving Hamiltonian \eqref{ham} can be compared  with recent experiments with $^{162}$Dy and $^{164}$Dy
making use of the following condition:
\begin{equation}
nr_0^3=\frac{na_d^3}{(a_d/r_0)^3}.
\end{equation}
As an example, setting an average density 
$n = 5 \cdot 10^{20} m^{-3}$, and with $n a^3_d = 2 \cdot 10^{-4}$, taking $a_d = 130a_0$ \cite{Kadau2016,Schmitt2016}, 
we have shown in Ref.\cite{PhysRevLett.119.215302} that 
experimental results lie in the transition region (within the error bars) 
from superfluid to cluster phase by tuning the scattering length. 

\section{Results}\label{section4}

\subsection{Methodology}\label{methods}

The low-temperature equilibrium properties of the systems described via the Hamiltonian \eqref{ham} have been investigated 
employing first-principle computer simulations based on the worm algorithm in the continuous-space path integral representation \cite{PhysRevLett.96.070601},
which allows one to essentially obtain the exact thermodynamics properties of Bose systems. 
Over the last ten years this computational technique has been successfully tested on a large variety of bosonic systems, including, for instance, $^{4}$He \cite{Boninsegni2006pre}, Rydberg atoms \cite{Cinti:2014aa} as well as dipolar systems 
\cite{Cinti2015,PhysRevA.95.023622,PhysRevA.92.053625,PhysRevE.92.052307,Jain2011}.
For the details of the implementation of the worm algorithm the reader may refer to Boninsegni et al. \cite{PhysRevLett.96.070601}. 
Due to the intrinsic anisotropy of the potential \eqref{v2}, we have implemented in this work the
so-called \textit{primitive approximation} for the 
imaginary time propagator, which requires a greater number of time slices 
with respect to more complex propagators, such as the pair-product ansatz \cite{RevModPhys.67.279} or the fourth-order one \cite{Jang2001}.
All the results reported below are extrapolated to the limit of zero temperature.
In this work we study the equilibrium properties of  scaled average density $nr^3_0 <  0.1$ and different 
interaction strengths ${\cal V}_0$. We work with $N$ atoms in a cubic box of linear dimension $L$ (fixing the density $n=N/L^3$) and  
using periodic boundary conditions. We have performed simulations with $N$ between 100 and 400; in such a limit one can exclude finite-size issues. From these simulations we obtain density profiles,  radial correlation functions, and the superfluid fraction.
The limit of zero temperature is obtained lowering the temperature until observables (e.g. energy, or superfluid fraction) do not change on further decreasing $T$.

\begin{figure}[t]
\centering
\includegraphics[width=12cm]{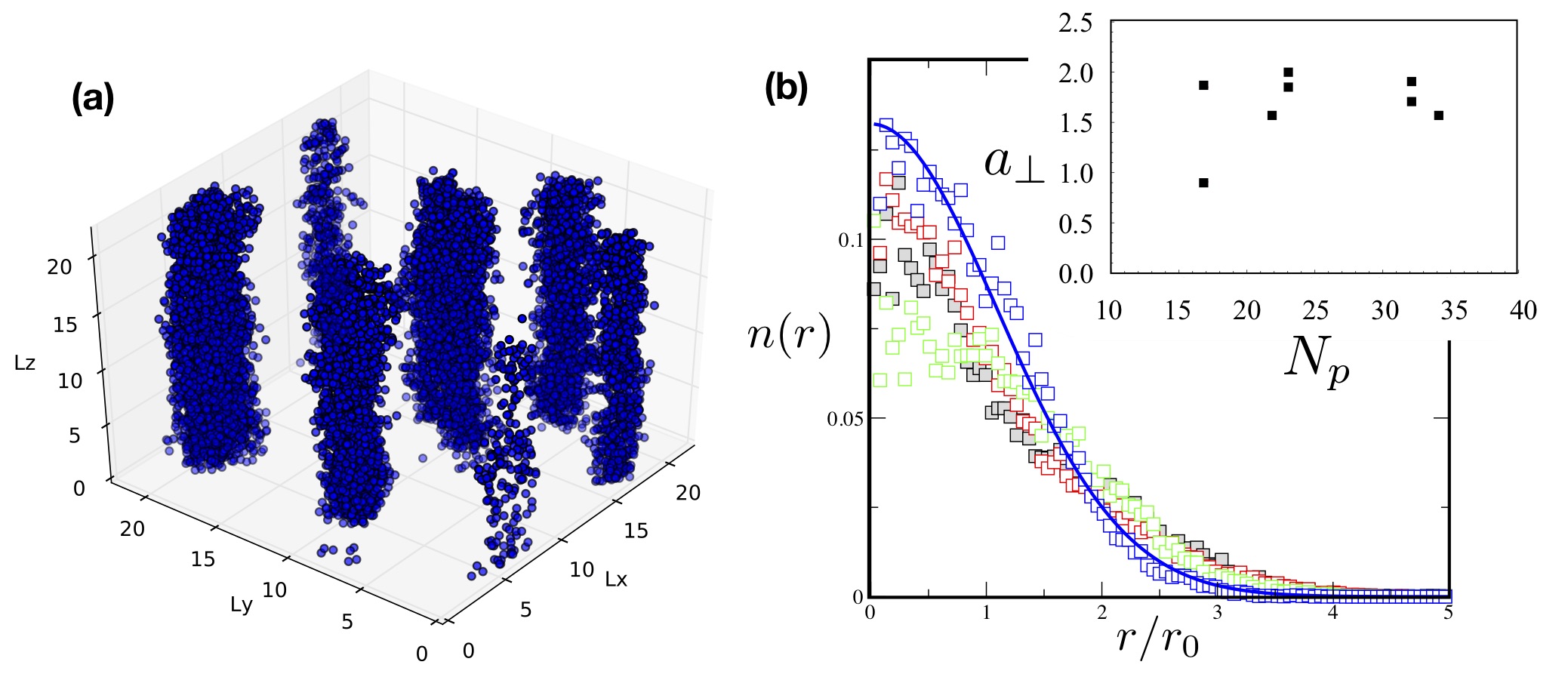}
\caption{\textbf{(a)} QMC density distribution of a filament phase for $n r_0^3 = 10^{-2}$. 
The number of filaments for these parameters is equal to four. Periodic boundary conditions have been employed.
\textbf{(b)} Particle density within a filament as a function of $r/r_0$. For these filaments the average particle number is about $N = 25$ (see text).}\label{figure3}
\end{figure}   

\subsection{Density profile of the Filament Phase}

In this section we present results applying the QMC method discussed in section \ref{methods}. 
We consider for now a regime of high density such as $n r_0^3 = 10^{-2}$. For this peculiar case 
our calculations obtained a stable filament phase at temperature much lower then the transition temperature 
from normal to superfluid liquid. Fig. \ref{figure3}a shows to a filament phase with $N=100$ and $P=1000$, $P$ being the number of imaginary-time slices. 
This snapshot displays the projection of the single-particle imaginary-time evolution onto the  real 
space. This representation of the condensate provides information of the delocalisation of particles in the box, and therefore of 
their probability distribution. 

The system shapes in four different well defined filaments, that we now characterize.
We fit the single-filament density with 
a gaussian $n_G(r)$
\begin{equation}
n_G(r) = \frac{N_p}{ \pi\, a_\perp^2\, L} e^{-\frac{r^2}{a_\perp^2}}
\end{equation}
 and extract the average number of particles within each filament. 
The inset in Fig.~\ref{figure3}b shows the width of the gaussian $a_\perp$ in units of $r_0$,  as a function of the particle number $N_p$. 
The estimate of the average inter-particle distance with a gaussian wave packet in the radial direction and a cylinder of length $L$ along $z$ yields
\begin{equation}
\left< r \right> \approx \frac{1}{n_f^{1/3}}= \left(\frac{\pi a_\perp^2 L}{N_p}\right)^{1/3} = 2.1\, r_0
\end{equation}
for the data in  Fig. \ref{figure3}a, where $n_f$ is the density of the filament. From the fit of the parameters above we get $N_p=25\pm7$ and $a_\perp/r_0= 1.7\pm0.4$.

\subsection{Weakly interacting regime}

With the many-body Hamiltonian \eqref{ham} one can also investigate the weakly interacting regime. 
To do that, we consider a limit where  $a_s$ and $a_d$ length scales are much smaller than the inter-particle
distance. This corresponds to maintain the general conditions $n|a_s|^3\ll1$ and $na_d^3\ll1$  simultaneously fulfilled. 
Quantitatively, as already discussed in Ref.\cite{PhysRevLett.119.215302}, both inequalities hold keeping $n|a_s|^3$ and $na_d^3 \leq5\cdot10^{-3}$. 

\begin{figure}[t!]
\centering
\includegraphics[width=15cm]{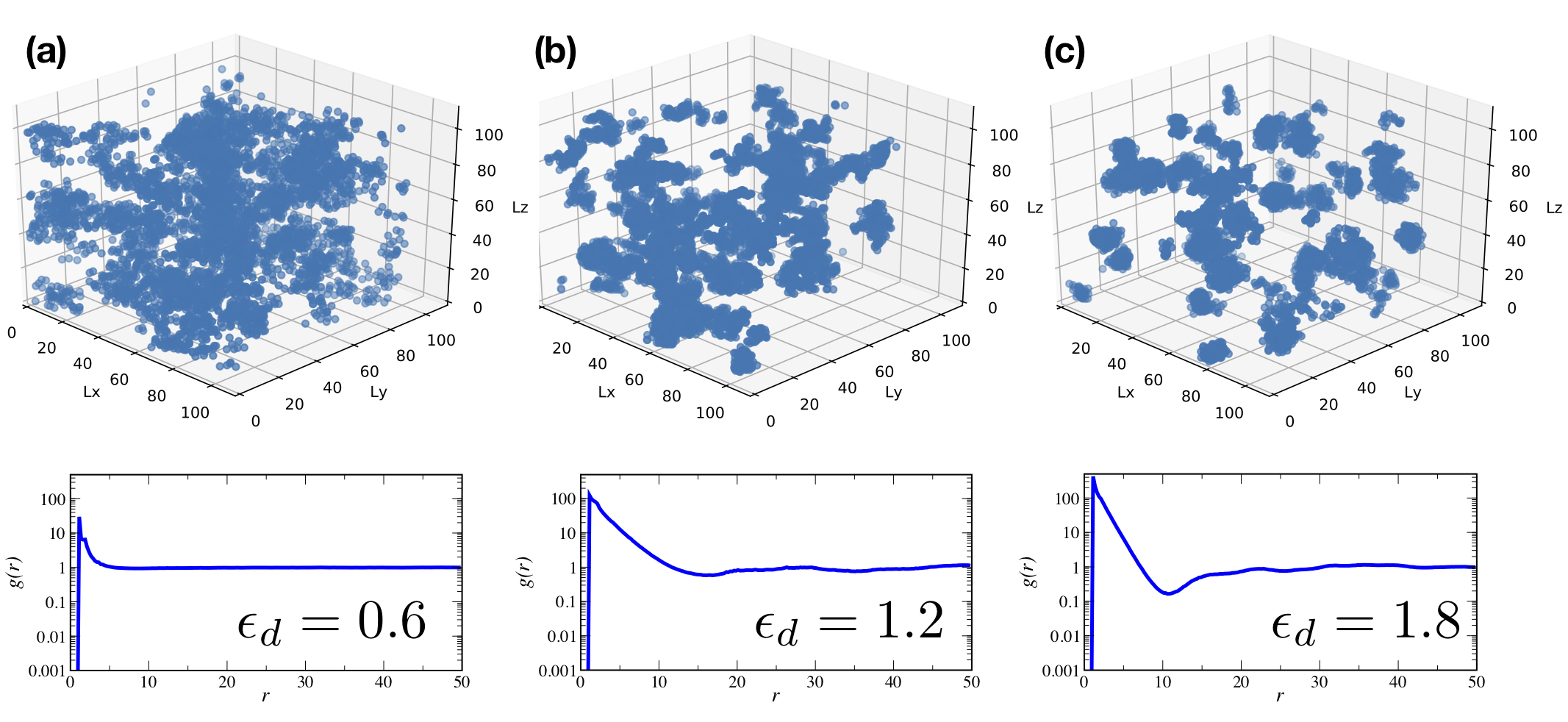}
\caption{Upper panels: QMC density distribution at different strengths of the dipolar interaction, for a rescaled density $n r_0^3 = 1\cdot10^{-4}$. Lower panels: radial distribution functions for the corresponding configurations.}\label{figure4}
\end{figure}   
We present QMC data at $\epsilon_d=0.6,\,1.2,\,1.8$. The upper part of Fig.~\ref{figure4} reports 
snapshot configurations at a rescaled low temperature equal to $T/T_0=0.25$, 
$T_0$ being the critical temperature of the ideal Bose gas $k_BT_0=2\pi(nr_0^3)^{2/3} / \zeta(3/2)^{2/3}$;
whereas the bottom panels show their corresponding radial distribution functions $g(r)$. 
Also in this case, simulations have been obtained with $N=100$, $P=1000$, keeping a fixed density of  $n r_0^3 = 1\cdot10^{-4}$. 

\begin{figure}[t]
\centering
\includegraphics[width=7.5cm]{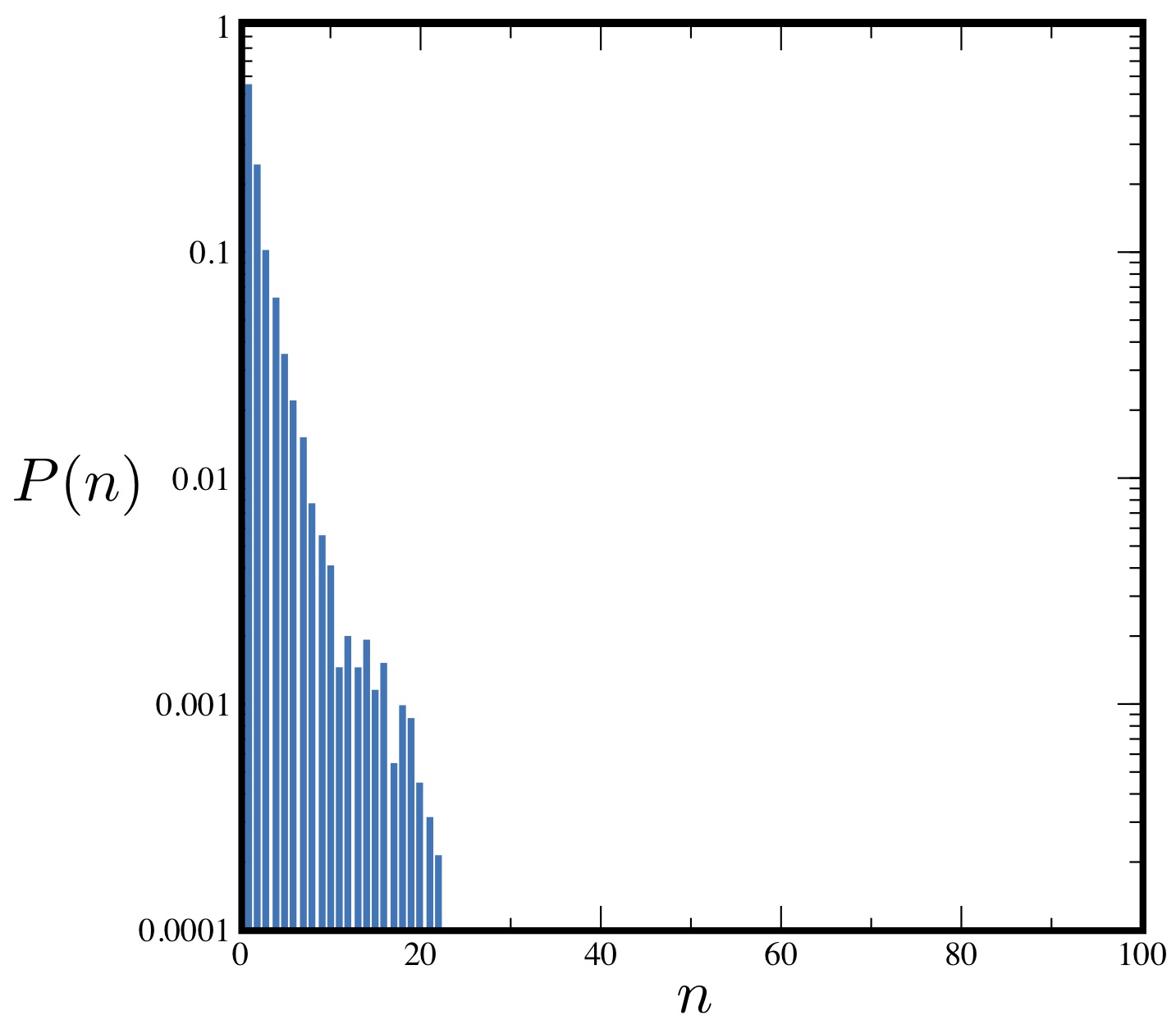}
\caption{Frequency of occurrence of cycles comprising $n$ particles at low $T/T_0$ (see text) for a system of $N = 100$ particles and $\epsilon_d=1.2$.}\label{figure5}
\end{figure}   
For $\epsilon_d=0.6$ (Fig.~\ref{figure4}a) the condensate appears modified by the dipolar strength only marginally. 
Radial distribution functions display strong correlations at short distances, 
presenting a typical fluid behavior at large $r$. 
Upon increasing the dipolar interaction, for $\epsilon_d>1$ (e.g. Fig.~\ref{figure4}b-c) 
we observe a stabilization of  clusters. As expected, their emergence exhibits a $g(r)$ with stronger correlations at 
short $r$ if compared to Fig.~\ref{figure4}a. 
Yet, the same observable in Fig.~\ref{figure4}b-c shows a gentle modulation at long distances, excluding long range crystalline order at larger $r$, and therefore any cluster crystal phase. Also, contrarily to the prediction of mean field theory for the simulation parameters, 
this phase is not globally superfluid.
In order to illustrate this point, we show the
computed the frequency $P(n)$ of the occurrence of permutation cycles involving $n$ particles ($1\leq n\leq N$) for the system with $\epsilon_d=1.2$  (see Fig.~\ref{figure5}). The important point here is that permutations extend only along cycles that remain confined within the single cluster, supporting the interpretation that locally particles exchange and quantum droplets are locally superfluid.
Finally, increasing further $\epsilon_d$ (Fig.~\ref{figure4}c), we observe nothing but the stabilization of the cluster phase. 
This completely excludes the existence of the filament regime discussed at higher densities.    

\section{Conclusion}\label{section5}


We discussed the appearance of cluster phases in ultracold dipolar bosons.
We reviewed the relevant pairwise interactions among dipoles, leading to the formation of clusters and filaments, as well as the numerical algorithm employed. We
characterized the many-body phases varying the strength of the potential as well as 
the density of the system via extensive QMC calculations. 
\funding{ T.M. acknowledges for support CAPES through the Project CAPES/Nuffic n.
88887.156521/2017-00 and CNPq through Bolsa de produtividade
em Pesquisa n. 311079/2015-6.}






\reftitle{References}





\end{document}